\documentclass[prb,amsfonts,amssymb,floats,twocolumn,aps]{revtex4-1}
\usepackage[utf8]{inputenc}
\usepackage[T1]{fontenc}
\usepackage{amsmath}
\usepackage{bbold}
\usepackage{float}
\usepackage{color}
\usepackage{braket}
\usepackage{bm}
\usepackage{graphicx}
\newcommand{\w}{\omega}
\newcommand{\Ztwo}{\mathbb{Z}_2}
\newcommand{\Uone}{\mathrm{U(1)}}
\newcommand{\SUtwo}{\mathrm{SU(2)}}

\newcommand{\HSW}{\mathcal{H}_{\rm SW}}
\newcommand{\Hzero}{\mathcal{H}_0}
\newcommand{\Htwo}{\mathcal{H}_2}

\graphicspath{{./}{figs/}}

\begin{document}

\title[]
{
Divergence of the Gr\"uneisen ratio at symmetry-enhanced first-order quantum phase transitions
}
\author{Charlotte Beneke}
\affiliation{Institut f\"ur Theoretische Physik and W\"urzburg-Dresden Cluster of Excellence ct.qmat, Technische Universit\"at Dresden,
01062 Dresden, Germany}
\author{Matthias Vojta}
\affiliation{Institut f\"ur Theoretische Physik and W\"urzburg-Dresden Cluster of Excellence ct.qmat, Technische Universit\"at Dresden,
01062 Dresden, Germany}

%%%%%%%%%%%%%%%%%%%%%%%%%%%%%%%%%%%%%%%%%%%%%%%%%%%%%%%%%%%%%%%%%%%%%%%%%

\date{\today}

\begin{abstract}
Studies of the Gr\"uneisen ratio, i.e., the ratio between thermal expansion and specific heat, have become a powerful tool in the context of quantum criticality, since it was shown theoretically that the Gr\"uneisen ratio displays characteristic power-law divergencies upon approaching the transition point of a continuous quantum phase transition. Here we show that the Gr\"uneisen ratio also diverges at a symmetry-enhanced first-order quantum phase transition, albeit with mean-field exponents, as the enhanced symmetry implies the vanishing of a mode gap which is finite away from the transition. We provide explicit results for simple pseudo-spin models, both with and without Goldstone modes in the stable phases, and discuss implications.
\end{abstract}


\maketitle

%%%%%%%%%%%%%%%%%%%%%%%%%%%%%%%%%%%%%%%%%%%%%%%%%%%%%%%%%%%%%%%%%%%%%%%%%
%%%%%%%%%%%%%%%%%%%%%%%%%%%%%%%%%%%%%%%%%%%%%%%%%%%%%%%%%%%%%%%%%%%%%%%%%
%%%%%%%%%%%%%%%%%%%%%%%%%%%%%%%%%%%%%%%%%%%%%%%%%%%%%%%%%%%%%%%%%%%%%%%%%

\section{Introduction}

Quantum phase transitions (QPTs) constitute an important topic in condensed-matter research:\cite{ssbook,mv_rop03,hvl} A QPT is associated with qualitative changes of the ground state of a many-body system, for instance its entanglement properties. Moreover, the quantum critical regime of continuous QPTs displays phenomenology very different from that of stable phases and is often the source for novel physics.
A number of signatures and tools have been identified to diagnose QPTs, such as critical power laws as function of absolute temperature $T$, with associated exponents, and universal scaling behavior.

Among the thermodynamic diagnostics for continuous QPTs, the Gr\"uneisen ratio (sometimes also called Gr\"uneisen parameter) -- defined as the ratio between thermal expansion $\alpha$ and specific heat $c_p$, $\Gamma = \alpha/c_p$ -- is particularly revealing:
In Ref.~\onlinecite{garst03} it was theoretically shown that it diverges upon approaching a pressure-driven quantum critical point (QCP) in a characteristic power-law manner, and also displays sign changes near a quantum critical point.\cite{garst05} For magnetic-field-driven transitions, the role of $\Gamma$ is taken by the magnetic Gr\"uneisen ratio, $\Gamma_H = -(\partial M/\partial T)_H/c_H$, which can be determined from the magnetocaloric effect.
Beyond standard quantum critical points, the Gr\"uneisen ratio has also been considered for disorder-dominated quantum Griffiths phases and found to display a much weaker logarithmic divergence as function of temperature.\cite{tvojta09}

On the experimental side, measurements of the Gr\"uneisen ratio have been frequently used to detect and characterize quantum phase transition. Prominent examples are the heavy-fermion compounds CeNi$_2$Ge$_2$,\cite{kuechler03} YbRh$_2$Si$_2$,\cite{kuechler03}, CeCu$_{6-x}$Ag$_x$ \cite{kuechler04}, and CeIn$_{3-x}$Sn$_x$ \cite{kuechler06} where a divergence of $\Gamma$ was found which could be attributed to a quantum phase transition.
Interestingly, divergencies of the magnetic Gr\"uneisen ratio have also been found in a number of compounds where no obvious quantum critical point exists, and we refer the reader to Ref.~\onlinecite{gegenwart17} for a review. At present, a consistent explanation for these observations is lacking.\cite{mv_rop18}

In the field of QPTs, a particularly interesting development concerns the emergence of enhanced symmetries at the transition point. These are symmetries not present in the underlying Hamiltonian, but emergent at long times and distances in the critical regime.
Such symmetries have been discussed in particular in the context of deconfined quantum critical points.\cite{senthil04} For instance, the transition between a N\'eel antiferromagnet and a columnar valence-bond solid\cite{sandvik07} has been argued to display an emergent SO(5) symmetry,\cite{nahum15} and a number of field-theoretic dualities have been invoked to rationalize enlarged symmetries.\cite{wang17} Likewise, enlarged symmetries have been detected in numerical simulations of $\Ztwo$ gauge theories coupled to Dirac fermions\cite{gazit18} and also at the ordering transition of a classical dimer model.\cite{sreejith19}

Symmetry enhancement is also possible at first-order QPTs. This refers to situations where the system discontinuously switches between two types of order, with the transition point displaying an emergent higher symmetry, leading to a family of stable states. Such behavior has been recently detected in numerical simulations of a SU(2)-symmetric spin model on the two-dimensional Shastry-Sutherland lattice where an emergent O(4) symmetry appeared at the transition between an antiferromagnet and a plaquette singlet state.\cite{sandvik19} In a related spin model on a square lattice, such a first-order transition has been found to display emergent SO(5) symmetry.\cite{sandvik20} Other examples of enhanced symmetries at first-order transitions appeared in Refs.~\onlinecite{serna19,yu19}. Together, these findings motivate to consider the phenomenology of symmetry-enhanced first-order QPTs in more detail, not the least to provide guidance to experiments.

In this paper we argue that the Gr\"uneisen ratio does not only diverge at quantum critical points, but generically also at symmetry-enhanced first-order QPTs. The reason is that the enhanced symmetry implies the existence of an excitation mode which is gapless only at the transition point but gapped away from it. We provide explicit results for simple effective pseudospin models for which we determine the full crossover behavior of the Gr\"uneisen ratio. We demonstrate that the Gr\"uneisen ratio displays not only a jump accompanied by a sign change upon crossing the transition at finite temperature, but also characteristic divergencies upon approaching the zero-temperature transition point. The type of divergencies depends on the presence or absence of Goldstone modes in the stable phases; the presence of Goldstone modes may lead to further sign changes of the Gr\"uneisen ratio. We also comment on further experimental implications.

The remainder of the paper is organized as follows: In Section~\ref{sec:gen} we summarize properties of the Gr\"uneisen ratio and argue why a divergence can be expected at symmetry-enhanced first-order QPTs. Section~\ref{sec:models} introduces the two spin models which we use to exemplify this divergence, with explicit results shown in Sections~\ref{sec:xz} and \ref{sec:xxz} for an Ising-Ising and a XY-Ising transition, respectively. A general discussion in Sec.~\ref{sec:concl} concludes the paper. Technical details are relegated to the Appendix.

%%%%%%%%%%%%%%%%%%%%%%%%%%%%%%%%%%%%%%%%%%%%%%%%%%%%%%%%%%%%%%%%%%%%%%%%%%%%%%%
%%%%%%%%%%%%%%%%%%%%%%%%%%%%%%%%%%%%%%%%%%%%%%%%%%%%%%%%%%%%%%%%%%%%%%%%%%%%%%%
%%%%%%%%%%%%%%%%%%%%%%%%%%%%%%%%%%%%%%%%%%%%%%%%%%%%%%%%%%%%%%%%%%%%%%%%%%%%%%%

\section{Gr\"uneisen ratio: General considerations}
\label{sec:gen}

A QPT occurs at $T=0$ as function of a non-thermal control parameter such as pressure, magnetic field, or chemical substitution. In the pressure-driven case, with the transition located at $p_c$, one may define a dimensionless control parameter $r=(p-p_c)/p_c$. At a continuous QPT, the free-energy density contains a critical contribution which can be probed via the volume thermal expansion
\begin{equation}
\label{alphadef}
\alpha=\frac{1}{V}  \left.\frac{\partial V}{\partial T}\right|_{p}=
-\frac{1}{V}   \left.\frac{\partial S}{\partial p}\right|_{T} \,.
\end{equation}
and the specific heat capacity
\begin{equation}
\label{cdef}
c_p=\frac{T}{N} \left.\frac{\partial S}{\partial T}\right|_{p} \,.
\end{equation}
where $S$ is the entropy, $V$ the sample volume, and $N$ the particle number.
The Gr\"uneisen ratio is commonly defined as\cite{gruendef}
\begin{eqnarray}
\label{gammadef}
\Gamma=\frac{\alpha}{c_p} =
-\frac{1}{V_n T}\frac{(\partial S/\partial p)_T}{(\partial S/\partial T)_p}
\end{eqnarray}
where $V_n=V/N$ the volume per particle.
For field-driven transitions, the dimensionless control parameter is $r=(H-H_c)/H_c$, such that one can define a quantity $\Gamma_H = (1/T) (\partial S/\partial H)_T / (\partial S/\partial T)_H$, also known as magnetic Gr\"uneisen ratio, which takes the role of $\Gamma$.

Using thermodynamic hyperscaling arguments, Ref.~\onlinecite{garst03} showed that the Gr\"uneisen ratio diverges in the quantum critical regime of a continuous QPT according to
\begin{equation}
\label{gr1}
\Gamma_{\rm cr}(T,r=0) \propto T^{-1/(\nu z)}
\end{equation}
where $\nu$ and $z$ are the correlation-length and dynamic exponents, respectively. Likewise, approaching the QCP at low temperatures yields the divergence
\begin{equation}
\Gamma_{\rm cr}(T=0,r) \propto |r|^{-1}\,.
\label{gr2}
\end{equation}
More precisely, Eqs.~\eqref{gr1} and \eqref{gr2} are obeyed in the regimes $T\gg |r|^{\nu z}$ and $T\ll |r|^{\nu z}$, respectively. Ref.~\onlinecite{garst03} also showed that these divergencies also hold (up to possible logarithmic corrections) for systems where hyperscaling does not apply, i.e., above the upper critical dimension.

Importantly, $\Gamma$ does not diverge upon approaching a finite-$T$ phase transition (although it can be enhanced near finite-temperature critical endpoints\cite{bartosch10,souza15,souza20}), thus a divergence of $\Gamma$ is usually considered a unique signature of a continuous QPT. Notably, for self-dual QCPs it can be shown that the Gr\"uneisen ratio remains finite, as the prefactor of the leading divergence vanishes;\cite{garst05,zhang19} this applies for instance to the transverse-field Ising model in one space dimension.

The scaling arguments put forward in Ref.~\onlinecite{garst03} describe the vicinity of a QCP, but more generally are valid if an excitation mode becomes soft as a function of a non-thermal control parameter. This is precisely what also happens at a symmetry-enhanced first-order QPT: The enhanced symmetry implies a larger number of soft modes at the transition point (as compared to away from it), hence we expect a divergent Gr\"uneisen ratio. This will be demonstrated explicitly in the remainder of the paper.

%%%%%%%%%%%%%%%%%%%%%%%%%%%%%%%%%%%%%%%%%%%%%%%%%%%%%%%%%%%%%%%%%%%%%%%%%%%%%%%
%%%%%%%%%%%%%%%%%%%%%%%%%%%%%%%%%%%%%%%%%%%%%%%%%%%%%%%%%%%%%%%%%%%%%%%%%%%%%%%
%%%%%%%%%%%%%%%%%%%%%%%%%%%%%%%%%%%%%%%%%%%%%%%%%%%%%%%%%%%%%%%%%%%%%%%%%%%%%%%

\section{Effective spin models}
\label{sec:models}

The microscopic models for which non-trivial symmetry-enhanced first-order QPTs have been established are complicated and not amendable to simple approximate solutions. We will therefore analyse toy models where the enhanced symmetry is explicit instead of emergent; these models should be understood as \emph{effective} models describing the relevant degrees of freedom near a symmetry-enhanced first-order QPT.

Specifically, we will consider spin models with tunable magnetic anisotropy. In the context of the first-order QPTs of interest, the models' degrees of freedom are to be interpreted as effective (pseudo)spins, and the models' anisotropy, encoded in a tuning parameter $\lambda$, can be tuned by hydrostatic pressure (or a similar control parameter). Thus, the enhanced symmetry at the transition point does \emph{not} correspond to an enhanced explicit symmetry of the original system. This aspect will become relevant when interpreting the results for thermal expansion, and we will get back to this below.

\subsection{Models}

\begin{figure}[t]
\includegraphics[width=0.95\linewidth]{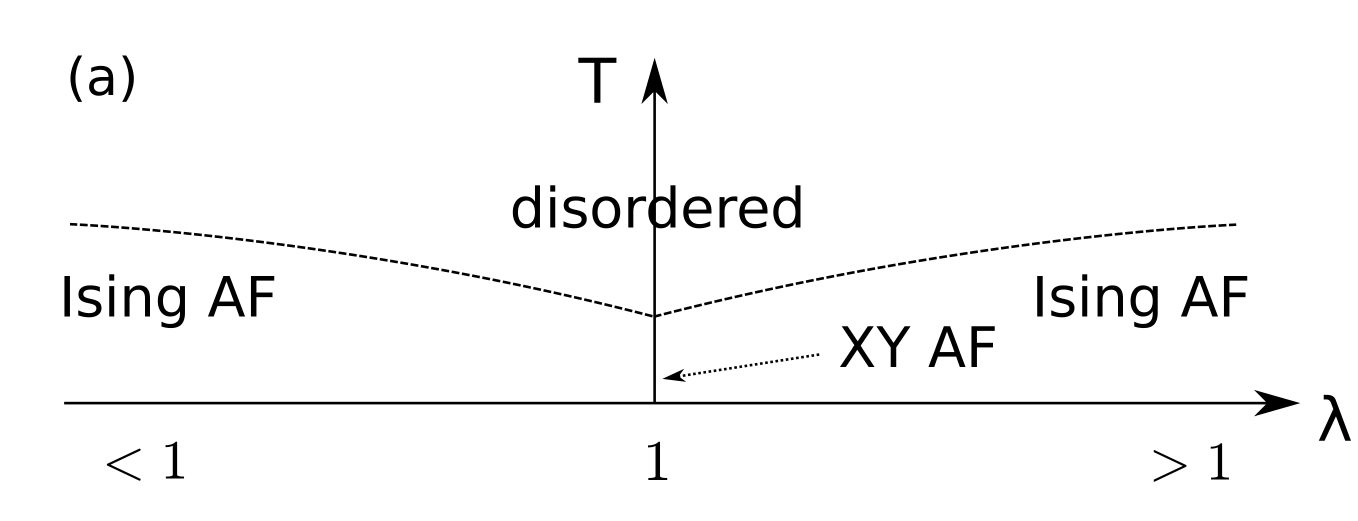}
\includegraphics[width=0.95\linewidth]{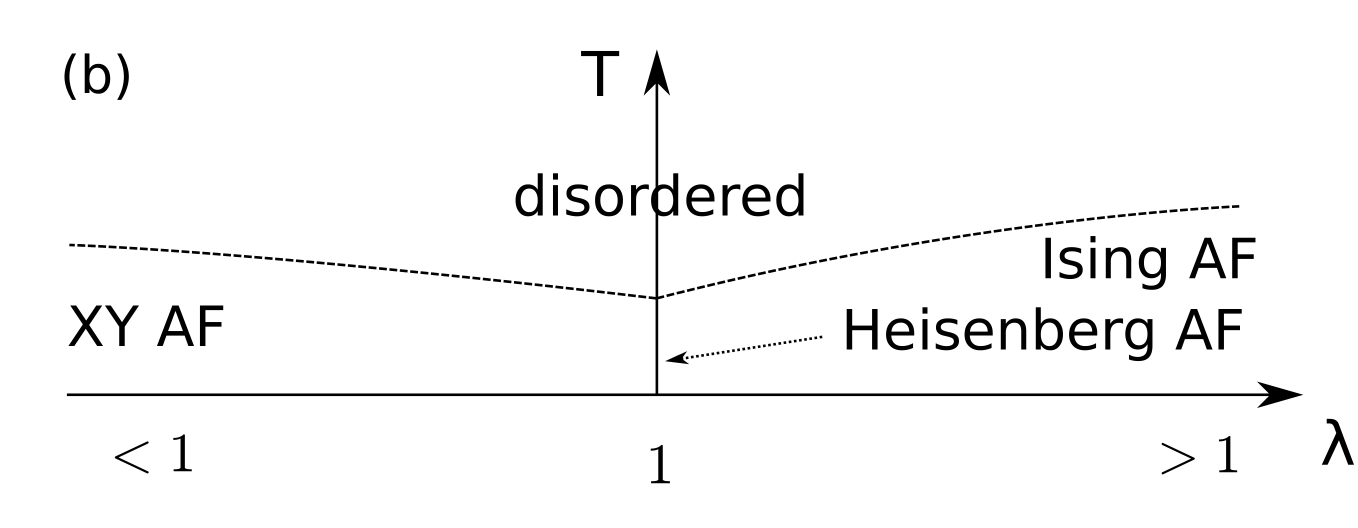}
\caption{
Schematic phase diagrams of the (a) XZ and (b) XXZ models in space dimension $d>2$. A symmetry-enhanced first-order QPT occurs upon tuning $\lambda$ through $\lambda\!=\!1$.
}
\label{fig:schematicphasediagrams}
\end{figure}

We will study the thermodynamics of two simple nearest-neighbor lattice spin models which display symmetry-enhanced first-order transitions. The first, which we dub XZ model, is defined as
\begin{equation}
\label{hxz}
\mathcal{H}_{XZ} = \sum_{\langle ij\rangle} (J_x S_i^x S_j^x + J_z S_i^z S_j^z),
\end{equation}
and the second is the XXZ model, with
\begin{equation}
\label{hxxz}
\mathcal{H}_{XXZ} = \sum_{\langle ij\rangle} [J_x (S_i^x S_j^x + S_i^y S_j^y) + J_z S_i^z S_j^z],
\end{equation}
where the ${\vec S}_i$ are spins of size $s$ located on sites $i$ of a regular lattice. For simplicity, we will work on a $d$-dimensional hypercubic lattice and consider $d=3$ unless noted otherwise.

In both models, we use $J_x\equiv J>0$ as the unit of energy, and parameterize the exchange anisotropy by $J_z=\lambda J_x$ with $\lambda>0$. At low temperatures, both models display antiferromagnetic long-range order: For $\lambda>1$ Ising  order is realized with spins along the $z$ direction. For $\lambda<1$ the XZ model displays Ising order along the $x$ direction whereas the XXZ model shows planar (XY-type) order. The XZ model is obviously symmetric (or self-dual) under the transformation $\lambda\to 1/\lambda$, $J\to\lambda J$ which exchanges the role of $x$ and $z$ directions in spin space. The key thermodynamic difference between both models is the existence of a Goldstone mode in the XXZ model for $\lambda<1$, whereas the XZ model is gapped for any $\lambda\neq 1$.

The point $\lambda=1$ displays enhanced $\Uone$ ($\SUtwo$) symmetry in the XZ (XXZ) model, respectively, and varying $\lambda$ through $1$ constitutes a symmetry-enhanced first-order transition. This implies that the order parameters of the phases realized for $\lambda\gtrless 1$ jump discontinuously if $\lambda$ is varied through 1. For instance, for the XZ model at $T=0$ and in the classical limit $s\to\infty$, the staggered magnetizations along the $x$ and $z$ directions follow $m_x = s \Theta(1-\lambda)$ and $m_z = s \Theta(\lambda-1)$, respectively, where $\Theta$ is the heaviside function.
Importantly, there is no simple phase coexistence or hysteresis associated with the transition at $\lambda=1$. Instead, the system displays order which spontaneously breaks a symmetry higher than in the adjacent $\lambda\gtrless 1$ phases. Specifically, at $\lambda=1$ the XZ (XXZ) model features one (two) Goldstone modes, respectively.
We recall that phase coexistence near conventional first-order transitions arises from the \emph{local} stability of both phases on both sides of the transition. The absence of phase coexistence is therefore a common property of symmetry-enhanced first-order transitions, as both phases do become locally unstable upon approaching the transition because they give way to the symmetry-enhanced manifold of states.

The finite-temperature phase diagrams are schematically shown in Fig.~\ref{fig:schematicphasediagrams}. The QPT continues as a vertical line of first-order transitions; in general such a line can be curved, but must display infinite slope as $T\to 0$ because the low-$T$ entropies on both sides of the transition are equal due to symmetry enhancement, $\lim_{T\to 0}\lim_{\lambda\to 1^-} S = \lim_{T\to 0}\lim_{\lambda\to 1^+} S$.

We assume that the anisotropy parameter $\lambda$ can be tuned by pressure: A change of external pressure $p$ changes the sample volume $V$, and the resulting changes of bond lengths lead to a change in $\lambda$, in other words, a volume change leads to a change of the effective anisotropy. Hence, $\partial S/\partial p = (\partial S/\partial \lambda) (\partial \lambda/\partial p)$, and the factor $(\partial \lambda/\partial p)$ is a system-specific constant.
We recall that the models are \emph{effective} models, hence the higher symmetry at $\lambda=1$ should not be confused with a higher explicit symmetry of the original system.

%%%%%%%%%%%%%%%%%%%%%%%%%%%%%%%%%%%%%%%%%%%%%%%%%%%%%%%%%%%%%%%%%%%%%%%%%%%%%%%

\subsection{Calculation of thermodynamics}

In $d=3$ space dimensions, the ordering temperature $T_N$ is finite\cite{yosida} at (and near) the transition point $\lambda=1$, see Fig.~\ref{fig:schematicphasediagrams}. We work at temperatures below $T_N$, such that antiferromagnetic order is well-established, and compute the thermodynamic quantities using standard linear spin-wave theory.
Using the Holstein-Primakoff representation of the spins, the bilinear piece of the Hamiltonian can be diagonalized using Fourier- and Bogoliubov transformations to yield a system of non-interacting magnon modes,
\begin{equation}
\label{eq:hswmagnons}
\HSW = \sum_{\vec{k},i} \w_{\vec{k},i} \alpha_{\vec{k},i}^\dagger \alpha_{\vec{k},i} + {\rm const}
\end{equation}
where the momentum summation $\sum_{\vec k}$ runs over the antiferromagnetic Brillouin zone of the ordered state, and $i$ is a mode index, for details see the Appendix.
The mode energies $\w_{\vec{k},i}$ characterize a free Bose gas, and one can compute the specific heat from the entropy according to
\begin{equation}
\label{eq:cv}
c_v = T^{-2} \frac{1}{N_s} \sum_{\vec{k},i} \frac{\w_{\vec{k},i}^2}{4\sinh^2 \w_{\vec{k},i}/(2T)}
\end{equation}
where $N_s$ is the number of lattice sites, and we have set Boltzmann's constant $k_B=1$. Note that we compute the specific heat at constant volume and neglect the difference between $c_v$ and $c_p$ (which is small in solids\cite{cpcvnote}).
Similarly, the thermal expansion follows from
\begin{equation}
\label{eq:al}
\alpha = T^{-2} \frac{1}{N_s} \sum_{\vec{k},i} \frac{\w_{\vec{k},i} \,\partial \w_{\vec{k},i}/\partial \lambda}{4\sinh^2 \w_{\vec{k},i}/(2T)}\,.
\end{equation}
As explained above, we have replaced the pressure derivative $\partial/\partial p$ with a $\lambda$ derivative $\partial/\partial\lambda$, assuming that $\partial \lambda/\partial p={\rm const}$.
Given that, due to the gap closing, at least one mode energy varies in a non-analytic fashion with $\lambda$ upon crossing the QPT, we can expect that both $c_v$ and $\alpha$ are non-analytic as function of $\lambda$ at $\lambda=1$.

In three space dimensions and in the absence of frustration, the linear spin-wave approximation provides reliable results at $T=0$ even for small spin size $s$ including $s=1/2$. This remains true at finite temperature as long as thermal occupations remain small, i.e., for $T\ll T_N$, with the mean-field estimate for the N\'eel temperature $T_N$ being $2dJ s^2$. In particular, the qualitative low-energy behavior of the modes near the QPT is dictated by Goldstone's theorem and will thus not change upon including interactions beyond linear spin-wave theory. We note that, in the leading order of the spin-wave expansion, the excitation energies scale as $(Js)$ which thus sets the natural unit for temperature. In the following we restrict our attention to the regime $T/(Js) < 1$.

%%%%%%%%%%%%%%%%%%%%%%%%%%%%%%%%%%%%%%%%%%%%%%%%%%%%%%%%%%%%%%%%%%%%%%%%%%%%%%%

\section{Transition between two gapped phases}
\label{sec:xz}

We start by analyzing the XZ model \eqref{hxz} which displays a symmetry-enhanced transition between two gapped Ising phases. The qualitative behavior of specific heat, thermal expansion, and Gr\"uneisen ratio can be derived analytically in the regime $T\ll Js$. Here we summarize the results, with details given in the Appendix.

At the transition point, $\lambda=1$, the system is an antiferromagnet which spontaneously breaks $\Uone$ symmetry, with a single gapless magnon mode with linear dispersion. In the Ising phases realized for $\lambda\neq 1$ the mode gap scales as $\Delta \propto |\lambda-1|^{1/2}$.

We first focus on the high-temperature regime, $T\gg\Delta$. This includes $\lambda=1$ where a scaling analysis of Eq.~\eqref{eq:cv} yields $c_v \propto T^d$ which also holds for $\lambda\neq1$ as long as $T\gg\Delta$.
The behavior of the thermal expansion is determined by the mode evolution with $\lambda$. For $\lambda\neq1$ we have $\partial\Delta/\partial\lambda \propto \pm|\lambda-1|^{-1/2}$, with the two signs applying to $\lambda\gtrless 1$, respectively. A scaling analysis of the relevant integral \eqref{eq:al} in the limit $T\gg\Delta$ then yields $\alpha\propto \pm T^{d-2}$.
As a result, we obtain $\Gamma \propto \pm 1/T^2$. Remarkably, this power-law divergence agrees with the scaling result\cite{garst03} at a quantum critical point $\Gamma\propto 1/T^{1/\nu z}$ if we assume mean-field exponents $\nu=1/2$ and $z=1$. This underlines the common origin of the Gr\"uneisen divergence, namely a mode gap closing as $\Delta\propto|r|^{\nu z}$ where $r=\lambda-1$ in the present case.
We note that the divergence of the Gr\"uneisen ratio is not in conflict with the results of Ref.~\onlinecite{zhang19}: A divergent $\Gamma$ is forbidden for self-dual continuous quantum phase transitions, but not in the first-order case.

In the low-temperature regime, $T\ll\Delta$, both $c_v$ and $\alpha$ are exponentially small, and their ratio to leading order is given by $\Gamma = (1/\Delta) \partial\Delta/\partial\lambda$ which results in a divergence $\Gamma\propto 1/(\lambda-1)$. Again, this agrees with the scaling result at a quantum critical point.\cite{garst03} We note that the exponents of the Gr\"uneisen divergence are not expected to change upon inclusion of corrections to linear spin-wave theory, because the low-energy structure of the theory, and hence the character of the gap closing, is protected by Goldstone's theorem.

\begin{figure}
\includegraphics[width=\linewidth]{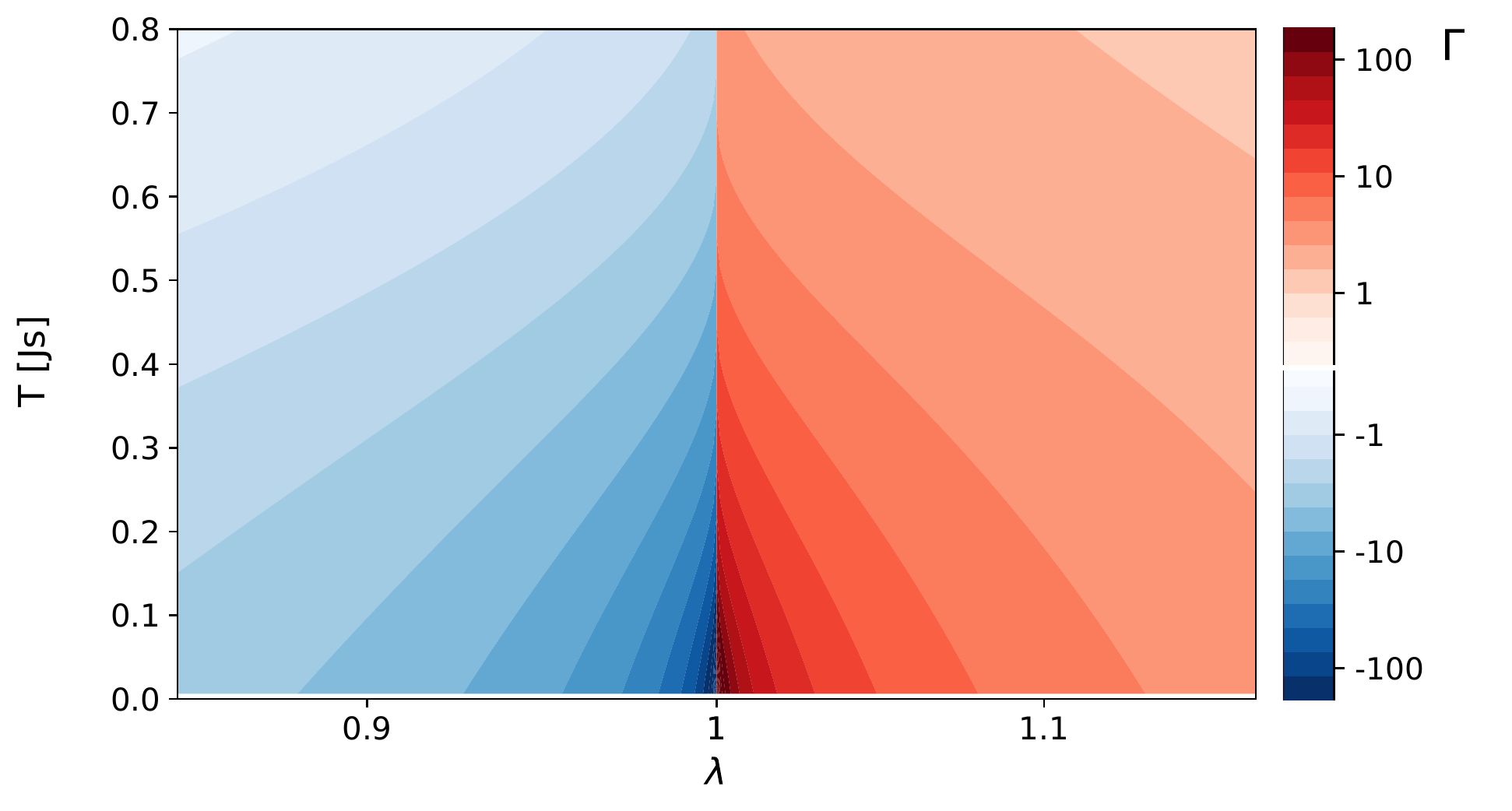}
\caption{
Gr\"uneisen ratio $\Gamma$ calculated for the XZ model \eqref{hxz} as function of tuning parameter $\lambda$ and temperature $T$. $\Gamma$ jumps and changes sign at the symmetry-enhanced first-order transition at $\lambda=1$ and diverges as $T\to 0$ both for $\lambda\to 1^+$ and $\lambda\to 1^-$.
Note that the $\lambda$ axis is logarithmic, emphasizing the self-duality\cite{sympicnote} of the model w.r.t. $\lambda\leftrightarrow 1/\lambda$.
}
\label{fig:xz_gruen1}
\end{figure}

\begin{figure}
\includegraphics[width=0.98\linewidth]{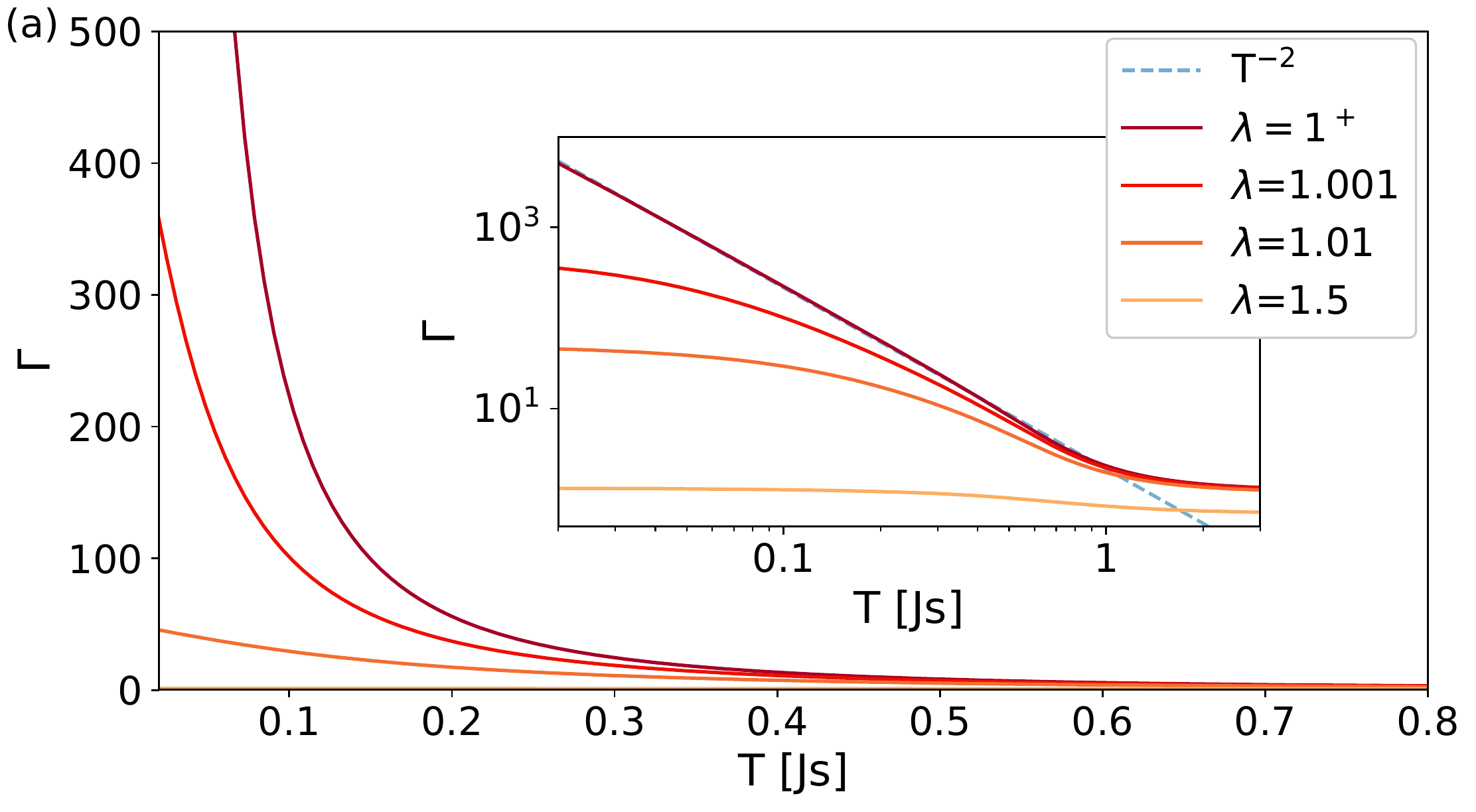}
\includegraphics[width=\linewidth]{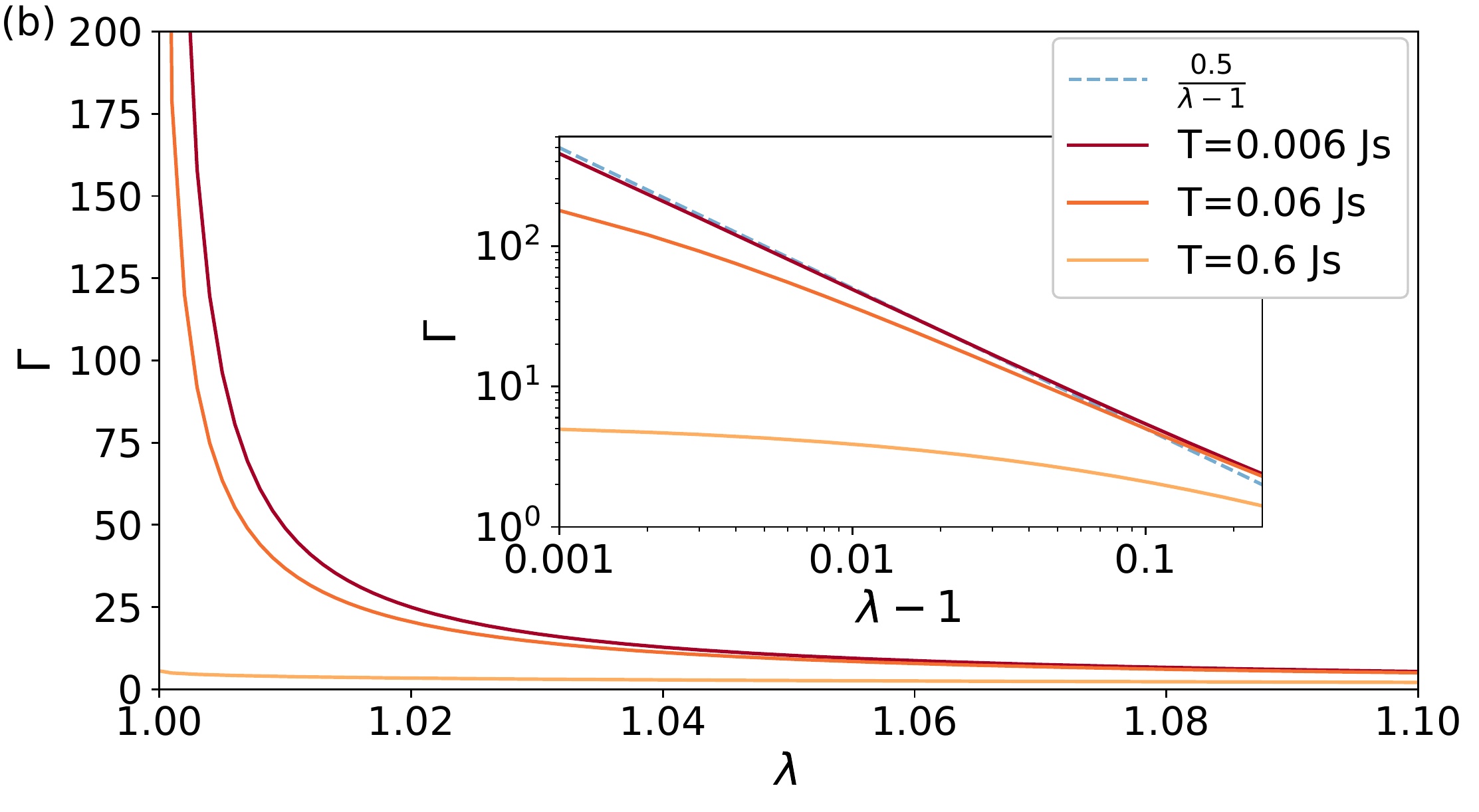}
\caption{
Gr\"uneisen ratio $\Gamma$ calculated for the XZ model \eqref{hxz} as in Fig.~\ref{fig:xz_gruen1}, here plotted (a) as function of $T$ for various $\lambda>1$ and (b) as function of $\lambda$ for various $T$. The insets show the data in a log-log fashion to illustrate the power laws.
}
\label{fig:xz_gruen2}
\end{figure}

The self-duality of the model implies \cite{sympicnote} $\Gamma(\lambda, T) = -\Gamma(1/\lambda, \lambda T)$ for small $|\lambda-1|$ and low $T$. This is consistent with $\Gamma \gtrless 0$ for $\lambda\gtrless 1$, i.e., $\Gamma$  jumps from negative to positive values upon crossing the transition at any $T$. This is different from the behavior in the quantum critical regime of a continuous QPT where $\Gamma$ varies analytically at finite $T$ which implies that $\Gamma$ displays a characteristic zero crossing within the quantum critical regime.\cite{garst05}

The analytical considerations are well borne out by our numerical calculations. Numerical results for the Gr\"uneisen ratio $\Gamma$, obtained from a lattice evaluation of Eqs.~\eqref{eq:cv} and \eqref{eq:al}, are displayed in Figs.~\ref{fig:xz_gruen1} and \ref{fig:xz_gruen2}. They show the full crossover behavior in the vicinity of the first-order QPT.

We finally comment on the behavior \textit{at} $\lambda=1$. Here, both thermal expansion and Gr\"uneisen ratio jump and change sign, i.e., are not well defined.
%\todo{check that cv is really the same, also at the moment the values of cv and alpha must be multiplied by 2 to fit the definition with $1/N_s$}.
This is a result of the assumption that a volume change controls the tuning parameter $\lambda$, which implies that a system placed at $\lambda=1$ will change its $\lambda$ value if heated or cooled at fixed $p$. In other words, changing $T$ at fixed $p$, starting at $\lambda=1$, drives the system in one of the stable phases.
Again, we recall that the enhanced $\Uone$ symmetry of our model at $\lambda=1$ is an emergent symmetry of the original system, being realized only on a particular line in $p$--$T$ parameter space.

%%%%%%%%%%%%%%%%%%%%%%%%%%%%%%%%%%%%%%%%%%%%%%%%%%%%%%%%%%%%%%%%%%%%%%%%%%%%%%%

\section{Transition between gapless and gapped phases}
\label{sec:xxz}

We now turn to the XXZ model whose main difference w.r.t. the XZ model is the Goldstone mode existing for $\lambda<1$ due to the spontaneously broken $\Uone$ symmetry. As above, we use simple analytical considerations to determine the asymptotic thermodynamic behavior in the various regimes.

We start with the Ising ordered phase: For $\lambda\geq 1$ the behavior of the XZ and XXZ model are very similar, i.e., gapless linearly dispersing modes at $\lambda=1$ and a gap scaling as $\Delta\propto(\lambda-1)^{1/2}$ for $\lambda>1$, but the number of low-energy modes is doubled in the XXZ model. Hence, for $\lambda>1$ we recover the results $\Gamma\propto 1/T^2$ for $T\gg\Delta$ and $\Gamma\propto 1/(\lambda-1)$ for $T\ll\Delta$.

\begin{figure}
\includegraphics[width=\linewidth]{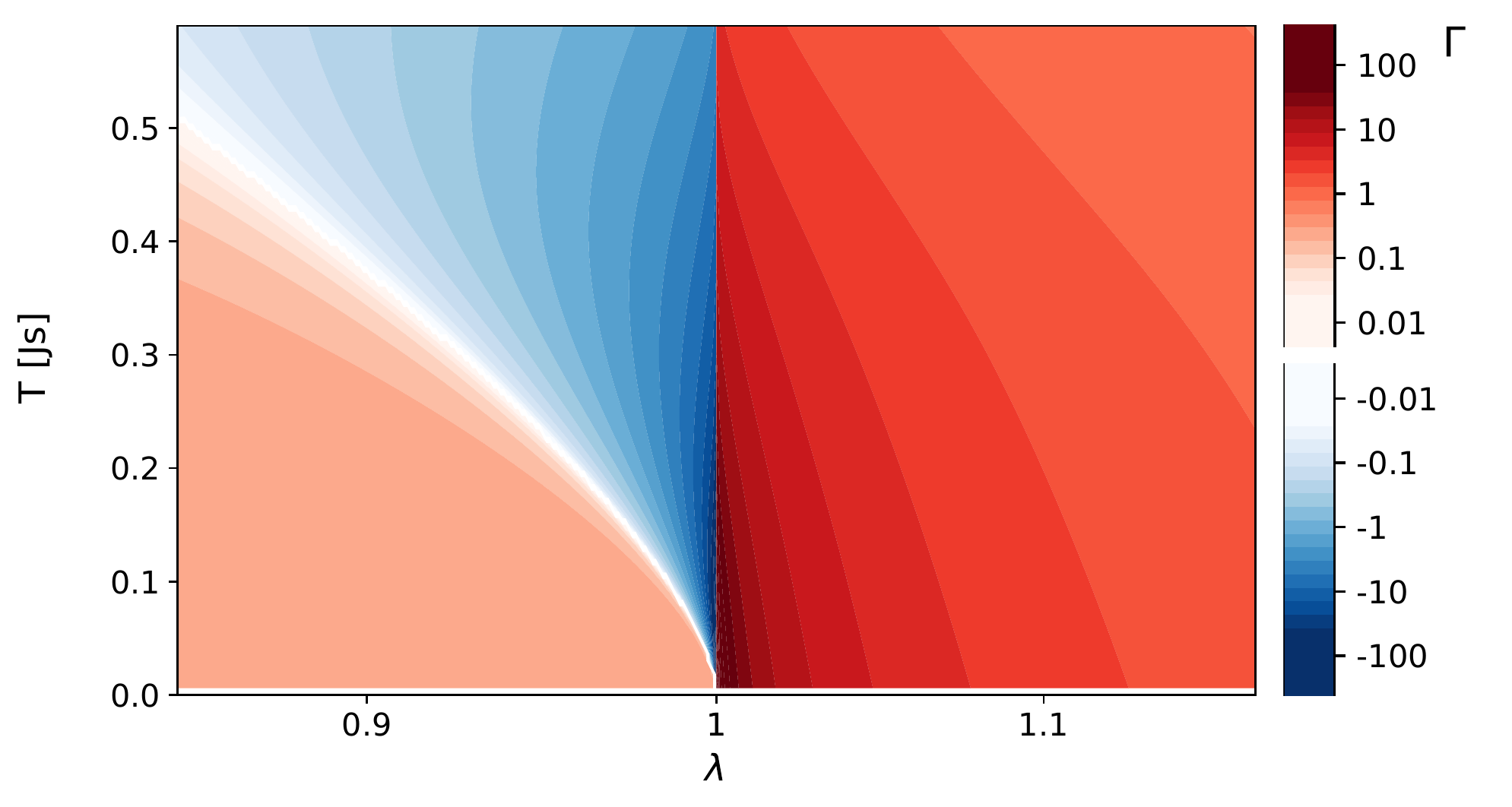}
\caption{
Gr\"uneisen ratio $\Gamma$ calculated for the XXZ model \eqref{hxxz} as function of tuning parameter $\lambda$ and temperature $T$. While $\Gamma$ diverges as $T\to 0$ for $\lambda\to 1^+$, the behavior for $\lambda\to 1^-$ is more complicated due to the presence of a Goldstone mode for $\lambda<1$, for details see text.
}
\label{fig:xxz_gruen1}
\end{figure}

\begin{figure}
\hspace*{2mm}%
\includegraphics[width=0.97\linewidth]{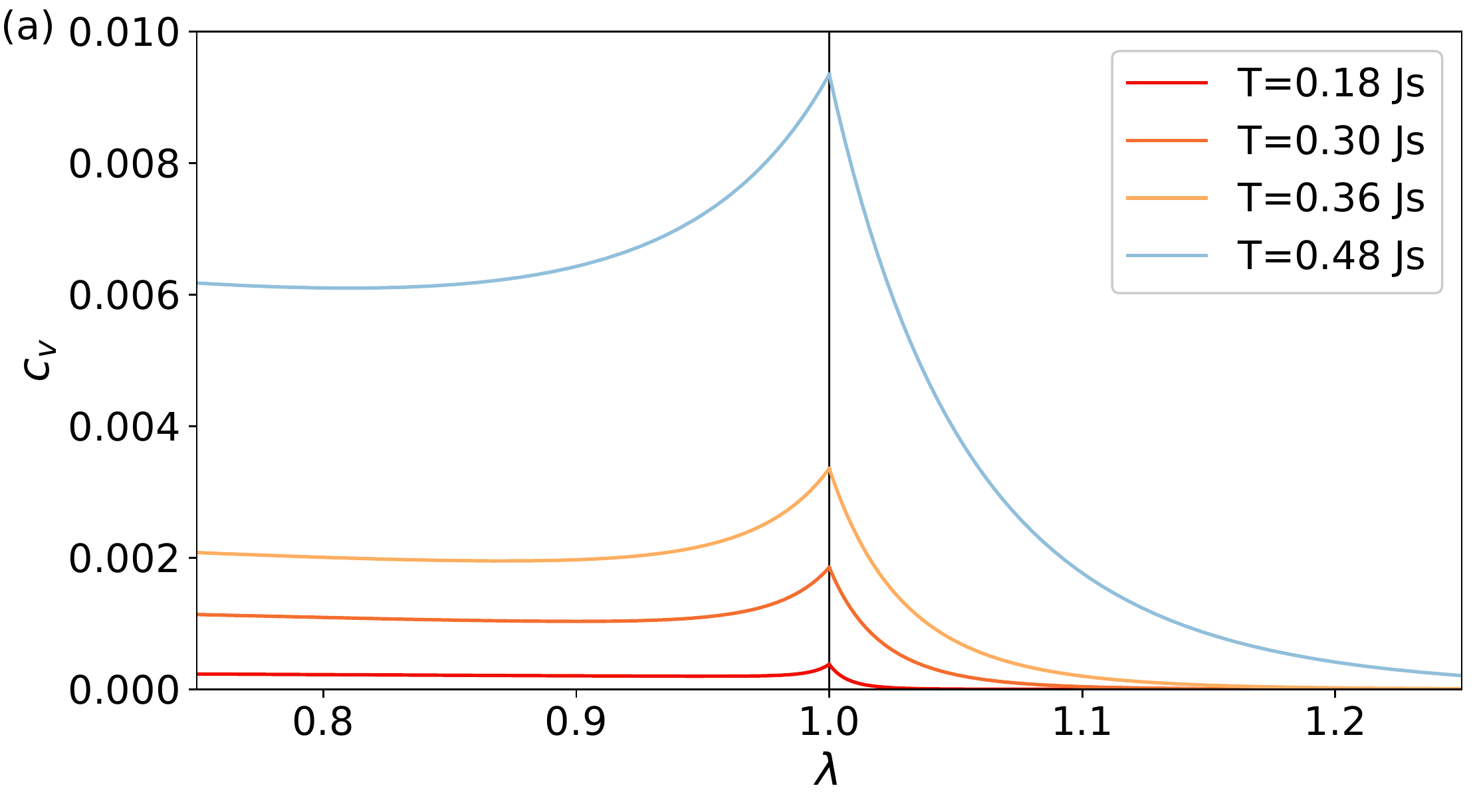}
\includegraphics[width=\linewidth]{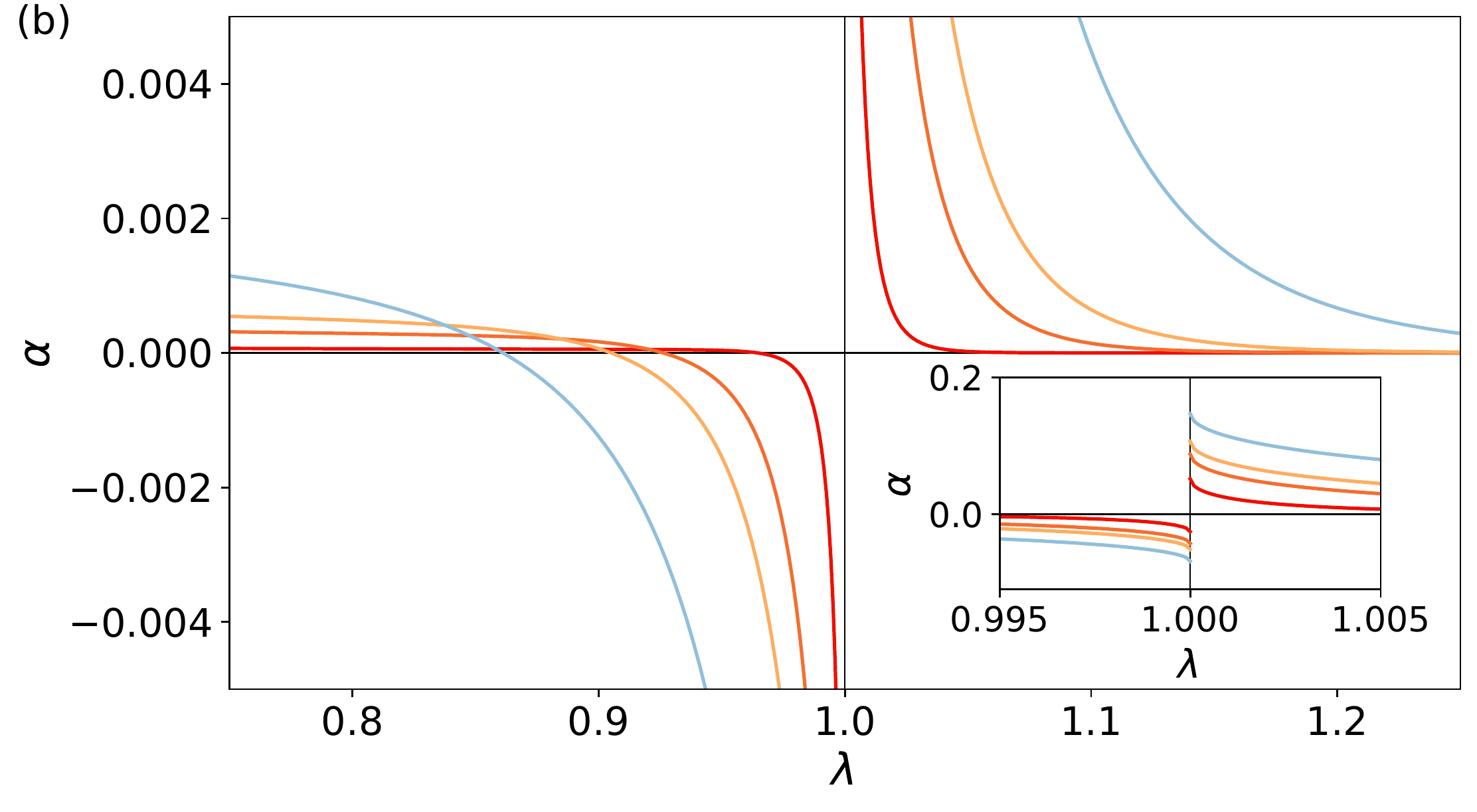}
\caption{
(a) Specific heat $c_v$ and (b) thermal expansion $\alpha$ calculated for the XXZ model \eqref{hxxz} as function of $\lambda$ for different $T$. $c_v$ is a continuous function of $\lambda$ while $\alpha$ jumps and changes sign at $\lambda=1$ as a result of the gap closing. A further sign change occurs for $\lambda<1$, for details see text.
}
\label{fig:xxz_cvalpha}
\end{figure}

\begin{figure}
\includegraphics[width=0.98\linewidth]{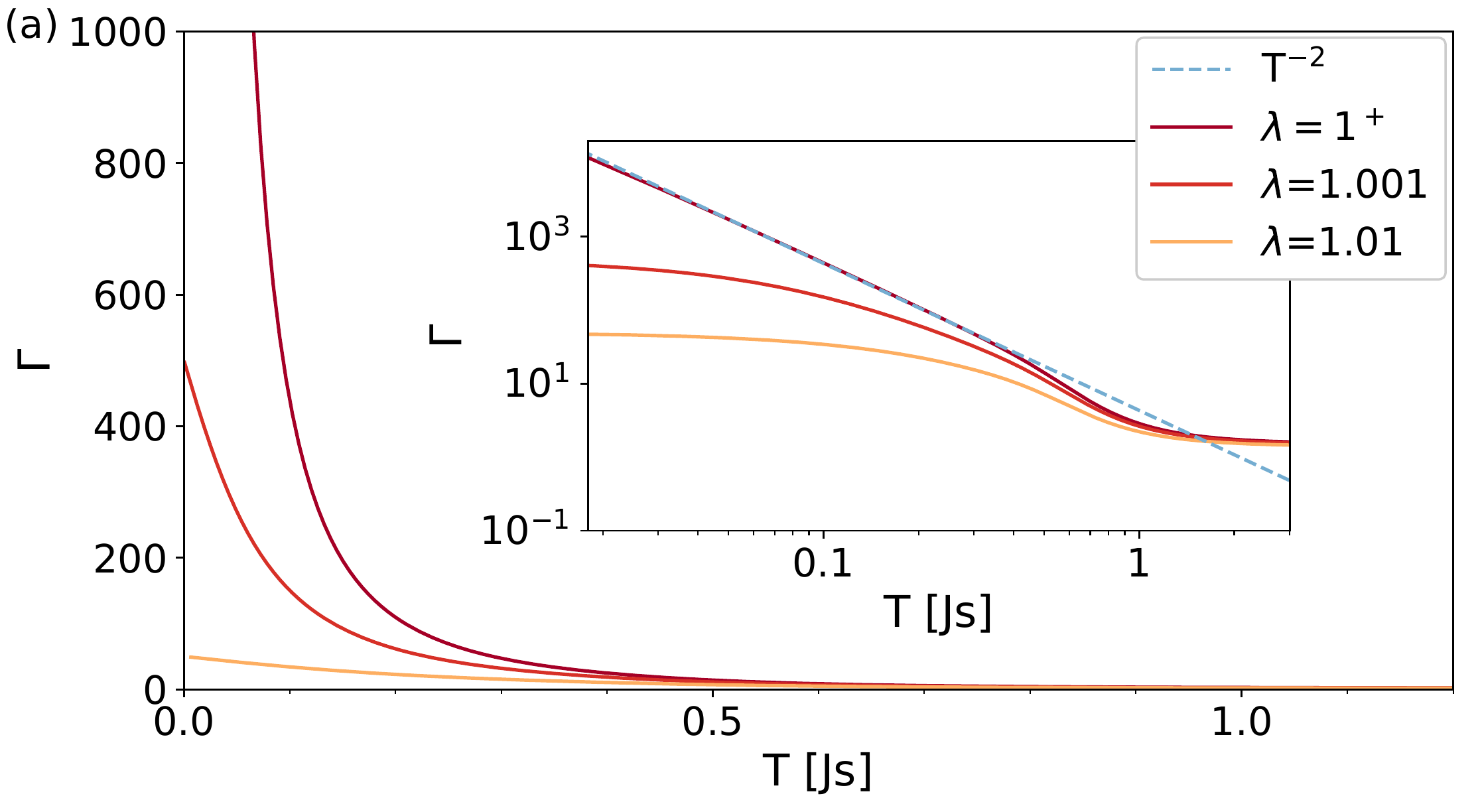}
\hspace*{2mm}%
\includegraphics[width=\linewidth]{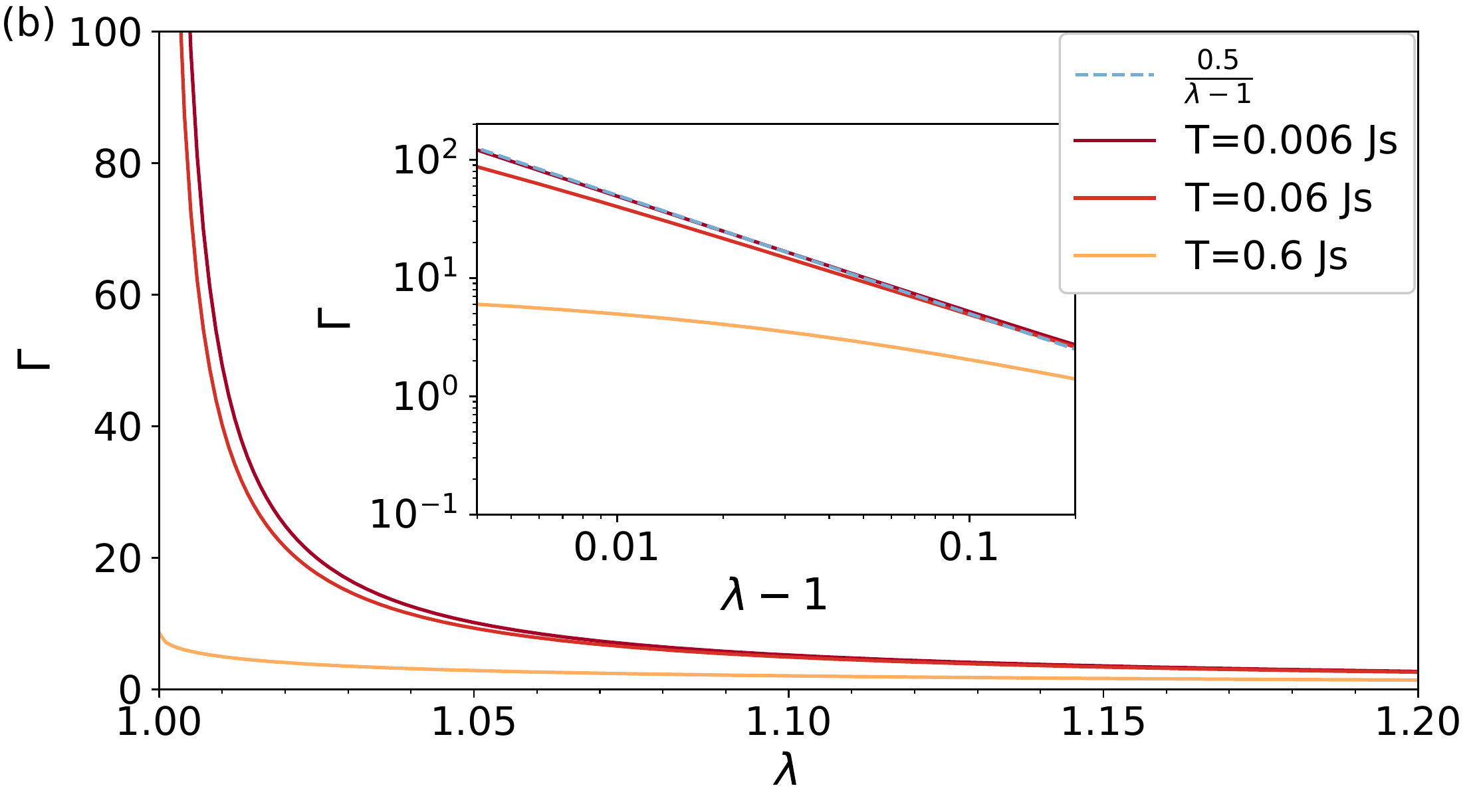}
\caption{
Gr\"uneisen ratio $\Gamma$ calculated for the XXZ model \eqref{hxxz} as in Fig.~\ref{fig:xxz_gruen1}, here plotted (a) as function of $T$ for various $\lambda>1$ and (b) as function of $\lambda>1$ for various $T$. The insets show the data in a log-log fashion to illustrate the power laws. In this gapped Ising phase, the behavior is very similar to that of the XZ model in Fig.~\ref{fig:xz_gruen2}.
}
\label{fig:xxz_gruen2}
\end{figure}

\begin{figure}
\includegraphics[width=\linewidth]{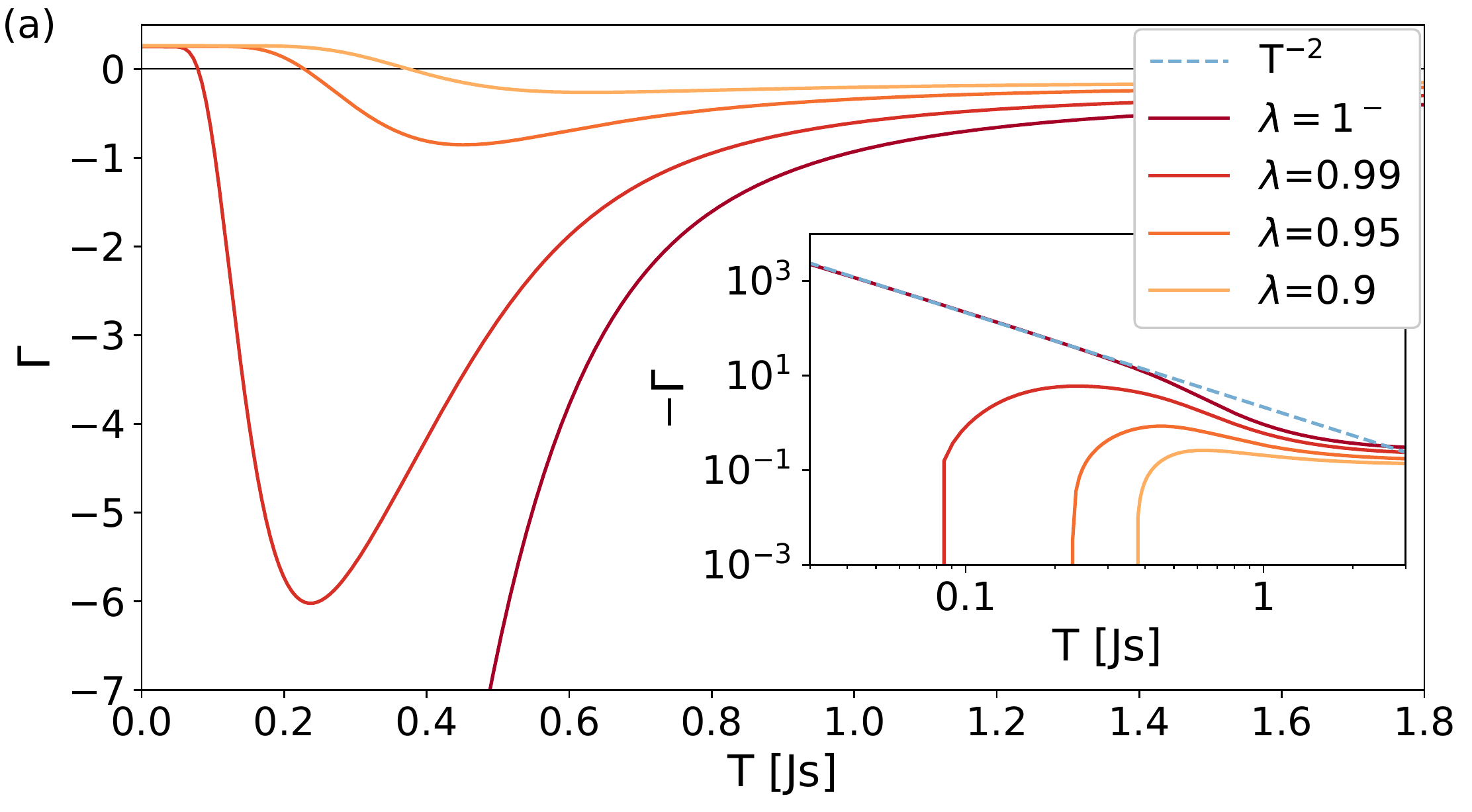}
\includegraphics[width=\linewidth]{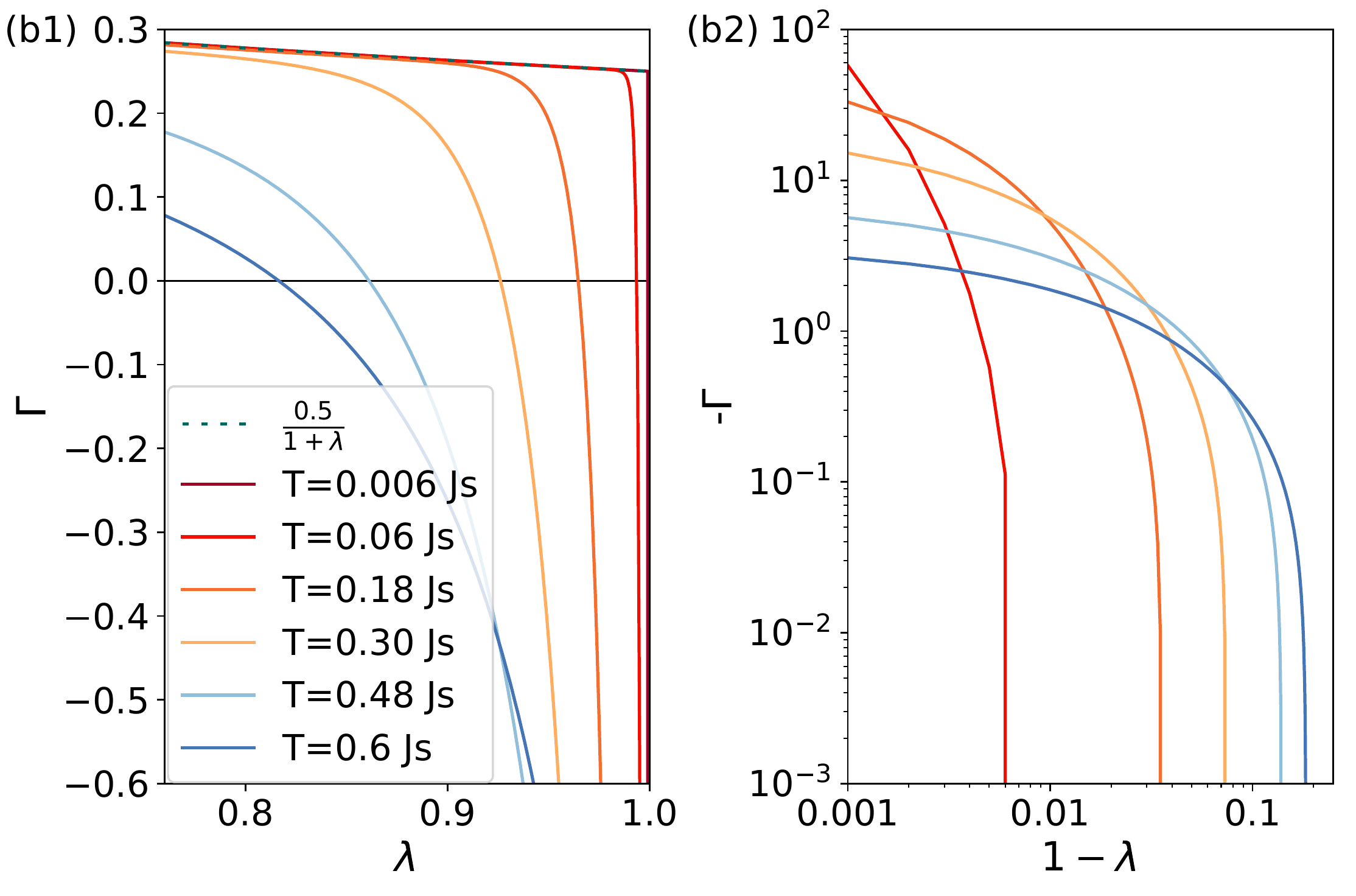}
\caption{
Gr\"uneisen ratio $\Gamma$ calculated for the XXZ model \eqref{hxxz} as in Fig.~\ref{fig:xxz_gruen2}, but now for the XY phase at $\lambda<1$. $\Gamma$ is shown (a) as function of $T$ for various $\lambda<1$ and (b1,b2) as function of $\lambda$ for various $T$.
}
\label{fig:xxz_gruen3}
\end{figure}

The situation is different for $\lambda<1$. Here, we have one Goldstone mode, corresponding to spin fluctuations in the XY plane, and one out-of-plane mode which develops a gap $\Delta\propto(1-\lambda)^{1/2}$. Within linear spin-wave theory, the mode contributions are strictly additive both for specific heat and thermal expansion, $c=c_{\Delta}+c_{gs}$ and $\alpha=\alpha_{\Delta}+\alpha_{gs}$.
The regime $T\gg\Delta$ has contributions from both modes. In this limit we find $\alpha_{gs}\propto T^d$ and $\alpha_\Delta\propto -T^{d-2}$, hence $|\alpha_{gs}|\ll|\alpha_{\Delta}|$ at low $T$, because the gapped mode displays a stronger  $\lambda$ dependence. Moreover, both $c_{\Delta}$ and $c_{gs}$ scale as $T^d$.
This results in a divergence $\Gamma\propto -1/T^2$, as in Section~\ref{sec:xz}.

The thermodynamics of the low-$T$ regime, $T\ll\Delta$, requires a more careful discussion. \emph{If} the contributions of the gapped out-of-plane mode are negligible compared to that of the Goldstone mode, we have to consider the latter only. As noted above, its contributions to both specific heat and thermal expansion scale as $T^d$, with prefactors both being non-singular as $\lambda\to 1^-$, see Appendix for details. Therefore, $\Gamma$ approaches a finite value in this case which, moreover, is positive, since the Goldstone-mode velocity increases with $\lambda$.
However, the condition $T\ll\Delta$ does \emph{not} automatically imply that the out-of-plane mode can be neglected, because in this regime we have $\alpha_\Delta \propto -T^{d/2-2} e^{-\Delta/T}$ while $\alpha_{gs} \propto T^{d}$. As a result, the Gr\"uneisen ratio changes sign for $\lambda<1$ along a line in the $T$-$\lambda$ phase diagram which ends at the $T=0$ transition point. The location of this line, given by $\alpha_{gs}+\alpha_\Delta=0$, can be estimated as $\Delta = (2+d/2) T |\ln T/(Js)|$ up to additive corrections, equivalently $1-\lambda \propto [T/(Js)]^2 \ln^2 T/(Js)$. Below this line, the Goldstone-mode contribution to $\alpha$ dominates and the Gr\"uneisen ratio is positive and finite, such that the behavior upon approaching the zero-temperature transition point is characterized by non-commuting limits,
\begin{equation}
\lim_{\lambda\to 1^-} \lim_{T\to 0} \Gamma(T,\lambda) \neq
\lim_{T\to 0} \lim_{\lambda\to 1^-} \Gamma(T,\lambda).
\end{equation}

The full numerical result for $\Gamma$ is displayed in Fig.~\ref{fig:xxz_gruen1}, with $c_v$ and $\alpha$ shown individually in Fig.~\ref{fig:xxz_cvalpha}. As anticipated, $\alpha$ and $\Gamma$ change sign twice. Further details of $\Gamma$ are in Figs.~\ref{fig:xxz_gruen2}, \ref{fig:xxz_gruen3}.
Its non-trivial behavior for $\lambda<1$, shown in Fig.~\ref{fig:xxz_gruen3}, becomes clear if one approaches the QPT along different trajectories. For square-root trajectories, i.e., fixed $\kappa=T/\Delta$, the Gr\"uneisen ratio diverges, $\Gamma(\kappa\Delta(\lambda),\lambda) \to -T^{-2}$ at sufficiently low $T$, as analytically shown in the Appendix. In contrast, along straight trajectories with fixed $\kappa' = T/[Js(1-\lambda)]$ the Gr\"uneisen ratio approaches a constant value, $\Gamma(\kappa'(1-\lambda)Js,\lambda) \to {\rm const.}$, as this trajectory is located below the line with $\Gamma=0$ sufficiently close to the QPT.

%%%%%%%%%%%%%%%%%%%%%%%%%%%%%%%%%%%%%%%%%%%%%%%%%%%%%%%%%%%%%%%%%%%%%%%%%%%%%%%
%%%%%%%%%%%%%%%%%%%%%%%%%%%%%%%%%%%%%%%%%%%%%%%%%%%%%%%%%%%%%%%%%%%%%%%%%%%%%%%
%%%%%%%%%%%%%%%%%%%%%%%%%%%%%%%%%%%%%%%%%%%%%%%%%%%%%%%%%%%%%%%%%%%%%%%%%%%%%%%

\section{Conclusions}
\label{sec:concl}

In this paper we have shown that the Gr\"uneisen ratio $\Gamma$, i.e., the ratio between thermal expansion and specific heat, generically diverges upon approaching a symmetry-enhanced first-order QPT, provided that it can be driven by pressure. Such a divergence, previously discussed and analyzed for continuous QPTs, occurs here because the enhanced symmetry is accompanied by a vanishing mode gap. Remarkably, the power laws characterizing the divergence of $\Gamma$ have the same form as found at continuous QPTs, for instance $\Gamma\propto \pm T^{-1/(\nu z)}$ asymptotically close to the transition, but with exponents locked to their mean-field values. This exponent locking is a result of the spontaneous symmetry breaking at the transition point, such that Goldstone's theorem protects the low-energy structure of the theory.

Our explicit results, obtained for ordered antiferromagnets, also demonstrate an interesting interplay of the soft mode(s) arising from symmetry enhancement with the Goldstone modes of the stable phases, leading to additional sign changes of $\Gamma$. Ordered phases with other symmetry breaking patterns can lead to different exponents for the Gr\"uneisen divergence; for example, a ferromagnetic version of the XZ model displays a divergence $\Gamma\propto \pm 1/T$ because $z\!=\!2$ in this case. It would be interesting to extend the present analysis to symmetry-enhanced first-order transitions without finite-temperature order where then non-mean-field exponents may be expected; this is beyond the scope of this paper.

The phenomenology outlined in this paper may be applicable to a number of correlated-electron materials where unconventional first-order-like transitions have been detected. One case in point is Ce$_3$Pd$_{20}$Si$_6$ where a field-driven switching between two magnetic phases, accompanied by a mode softening, has been observed in neutron scattering experiments.\cite{inosov19} In this context we note that symmetry-enhanced first-order transitions in metals will induce deviations from Fermi-liquid behavior. The details of this will be investigated in future work. Finally, we note that the closing of a mode gap may also occur in different first-order transition settings,\cite{oles07,delre16} and it would be interesting to extend the thermodynamic analysis presented here to those cases.

%%%%%%%%%%%%%%%%%%%%%%%%%%%%%%%%%%%%%%%%%%%%%%%%%%%%%%%%%%%%%%%%%%%%%%%%%%%%%%%

\acknowledgments

We thank P. M. Consoli, M. Garst, and L. Janssen for useful discussions as well as for collaborations on related subjects.
Financial support from the Deutsche Forschungsgemeinschaft through SFB 1143 (project ID 247310070) and the W\"urzburg-Dresden Cluster of Excellence \textit{ct.qmat} -- Complexity and Topology in Quantum Matter (EXC 2147, project ID 390858490) is gratefully acknowledged.

%%%%%%%%%%%%%%%%%%%%%%%%%%%%%%%%%%%%%%%%%%%%%%%%%%%%%%%%%%%%%%%%%%%%%%%%%%%%%%%
%%%%%%%%%%%%%%%%%%%%%%%%%%%%%%%%%%%%%%%%%%%%%%%%%%%%%%%%%%%%%%%%%%%%%%%%%%%%%%%
%%%%%%%%%%%%%%%%%%%%%%%%%%%%%%%%%%%%%%%%%%%%%%%%%%%%%%%%%%%%%%%%%%%%%%%%%%%%%%%

\appendix

\section{Details of spin-wave expansion}
%Write expressions which fix the Gr\"uneisen prefactors.

To determine the thermodynamics of the spin models \eqref{hxz} and \eqref{hxxz}, we employ standard linear spin-wave theory.\cite{yosida} In all phases, we expand about a two-sublattice N\'eel state and introduce two types of Holstein-Primakoff bosons $a$ and $b$ for the A and B sublattices, respectively. As a result, Fourier transformations are performed with momenta $\vec k$ from the magnetic Brillouin zone, with $N_s/2$ momentum points, where $N_s$ is the total number of lattice sites.

%%%%%%%%%%%%%%%%%%%%%%%%%%%%%%%%%%%%%%%%%%%%%%%%%%%%%%%%%%%%%%%%%%%%%%%%%%%%%%%

\subsection{XZ Model}

For the $\lambda\geq 1$ Ising phase of XZ model \eqref{hxz}, the spin-wave Hamiltonian reads $\HSW = \Hzero + \Htwo$, with $\Hzero = -2d \lambda J N_s s^2$ the classical ground-state energy and
\begin{align}
\Htwo
= 2d J s \sum_{\vec k} &\Big[ \lambda a^\dagger_{\vec{k}}a_{\vec{k}}+ \lambda b^\dagger_{\vec{k}}b_{\vec{k}} \notag\\
&+ \frac{\gamma_k}{2} (a^\dagger_{\vec{k}}b_{\vec{k}}+a_{\vec{k}}b_{-\vec{k}}+h.c.) \Big]
\end{align}
the bilinear fluctuation piece, with the form factor
\begin{align}
\gamma_k=\frac{1}{d}\sum_{j=1}^d \cos k_j
\label{Eq:gammak}
\end{align}
where we have set the lattice constant to unity. Note that $\gamma_k\geq0$ in the magnetic Brillouin zone.
$\Htwo$ is diagonalized by a standard $4\times4$ Bogoliubov transformation,\cite{wesselmilat} yielding two sets of eigenmodes $\alpha_{\vec k,i}$ with mode energies
\begin{align}
\omega_{\vec{k},1} &= 2d J s \sqrt{\lambda^2-\lambda\gamma_k}\,, \nonumber \\
\omega_{\vec{k},2} &= 2d J s \sqrt{\lambda^2+\lambda\gamma_k}\,.
\label{Eq.omega12xz}
\end{align}
At the $\Uone$-symmetric point, $\lambda=1$, $\omega_{\vec{k},1}$ represents the Goldstone mode of the system, with linear dispersion around $\vec k=0$. In contrast, $\omega_{\vec{k},2}$ is always gapped.

We proceed with an analytical calculation of the Gr\"uneisen ratio at low temperatures and near the transition point. In this regime, contributions of the mode $\omega_{\vec{k},1}$ are exponentially suppressed and can be neglected.
For small $k$ and small $(\lambda-1)$, $\omega_{\vec{k},1}$ can be expanded as
\begin{align}
\omega_{\vec{k},1} \approx \sqrt{\Delta^2_1 + c_1^2 |\vec{k}|^2}
\label{Eq:Taylorexcitationenergyspinmodel}
\end{align}
with the gap $\Delta_1\equiv \Delta$ given by
\begin{align}
\Delta_1(\lambda)=2dJs\sqrt{\lambda(\lambda-1)}
\label{eq:delta1}
\end{align}
and the velocity
\begin{align}
c_1(\lambda)=Js\sqrt{2d \lambda}.
\end{align}
To evaluate the thermal expansion, we need the $\lambda$ derivatives of gap and velocity,
\begin{align}
&\frac{\partial \Delta_1(\lambda)}{\partial \lambda}=2Jsd\frac{2\lambda-1}{2\sqrt{\lambda(\lambda-1)}} \nonumber \\ &\Rightarrow \quad
\Delta_1\frac{\partial \Delta_1(\lambda)}{\partial \lambda}=(2Jsd)^2\left(\lambda-\frac{1}{2}\right)
\end{align}
and
\begin{align}
\frac{\partial c_1(\lambda)}{\partial \lambda}=c_0 \frac{1}{2\sqrt{\lambda}}  \quad \Rightarrow \quad
\frac{1}{c_1(\lambda)}\frac{\partial c_1(\lambda)}{\partial \lambda}=\frac{1}{2\lambda}
\end{align}
with $c_0=c_1(1)$.

The expressions \eqref{eq:cv} and \eqref{eq:al} can now be evaluated in the continuum limit, making use of the spherical symmetry of the dispersion \eqref{Eq:Taylorexcitationenergyspinmodel}.
We start with the regime $T\gg\Delta$, reached, e.g., by taking the limit $\lambda\to 1^+$ at finite $T$. Using the substitution $x= \beta c_1(\lambda) k$ where $\beta=1/T$ we obtain
\begin{align}
&\lim_{\lambda\to 1^+} c_v
= C_1(1) \int_0^\infty \!\! dx \frac{x ^{d+1}}{4 \sinh^2 (x/2)}
\end{align}
and
\begin{align}
&\lim_{\lambda\to 1^+} \alpha
=C_1(1) \int_0^\infty \!\! dx \frac{\beta^2(2d Js)^2 x^{d-1}+ x^{d+1}}{8 \sinh^2 (x/2)}
\end{align}
with the prefactor
\begin{align}
C_1(\lambda)= \Omega_d \frac{T^d}{ c_1(\lambda)^d}
\label{Eq:Cl}
\end{align}
where $\Omega_d$ is defined as the $d$-dimensional solid angle of the hypersphere.
The leading low-$T$ behavior is thus $c_v\propto T^d$ and $\alpha\propto T^{d-2}$, with the dominant term in $\alpha$ arising from the $\lambda$ dependence of the gap. The integrals can be evaluated, resulting in
\begin{align}
\label{eq:gamxz0}
&\lim_{\lambda \rightarrow 1^+} \Gamma=
\frac{2 d\zeta(d\!-\!1)}{(d\!+\!1)\zeta(d\!+\!1)} \Big(\frac{T}{Js}\Big)^{-2}
\end{align}
where $\zeta(x)$ is the Riemann zeta function. The expression \eqref{eq:gamxz0} matches the numerical result in Fig.~\ref{fig:xz_gruen2}(a) for $d=3$ where the prefactor evaluates to $2.28$. Corrections to the leading power law \eqref{eq:gamxz0} take the form of a standard high-temperature expansion, $\Gamma \propto [T/(Js)]^{-2}[1+\mathcal{O}(\Delta/T)]$.

In the opposite limit $T \ll \Delta$, both specific heat and thermal expansion are exponentially suppressed. For $\lambda>1$, we have
\begin{align}
\lim_{T\to 0} c_v(\lambda)
&=C_1(\lambda) \ \beta^2\Delta_1^2 e^{-\beta\Delta_1} \nonumber \\
&\times \int_0^\infty dx \ x^{d-1}  e^{-x^2/(2\beta\Delta_1)}
\end{align}
and
\begin{align}
\lim_{T\to 0} \alpha(\lambda)
&=C_1(\lambda) \ \beta^2\Delta_1\frac{\partial \Delta_1}{\partial \lambda} e^{-\beta\Delta_1} \nonumber \\
&\times \int_0^\infty dx \ x^{d-1}  e^{-x^2/(2\beta\Delta_1)}.
\end{align}
The leading low-$T$ behavior of their ratio is thus found as
\begin{align}
\lim_{T\to 0} \Gamma(\lambda) = \frac{\frac{\partial \Delta_1(\lambda)}{\partial \lambda}}{\Delta_1(\lambda)} = \frac{2\lambda-1}{2\lambda(\lambda-1)} ~\underset{\lambda\to1^+}{\to}~ \frac{1/2}{\lambda-1}
\end{align}
in agreement with the numerical result in Fig.~\ref{fig:xz_gruen2}(b).

%%%%%%%%%%%%%%%%%%%%%%%%%%%%%%%%%%%%%%%%%%%%%%%%%%%%%%%%%%%%%%%%%%%%%%%%%%%%%%%

\subsection{XXZ Model}

In the XXZ model \eqref{hxxz} the low-temperature phase for $\lambda>1$ ($\lambda<1$) breaks a $\Ztwo$ ($\Uone$) symmetry, respectively. Therefore two different calculations are required.

\subsubsection{Ising Phase}

For $\lambda \geq 1$ the expansion is performed about a N\'eel state polarized along $\hat z$. The bilinear piece of the spin-wave Hamiltonian now reads
\begin{align}
\Htwo
= 2d J s \sum_{\vec k} &\Big[ \lambda a^\dagger_{\vec{k}}a_{\vec{k}}+ \lambda b^\dagger_{\vec{k}}b_{\vec{k}} + \gamma_k (a_{\vec{k}}b_{-\vec{k}}+h.c.) \Big]
\end{align}
with $\gamma_k$ defined in Eq.~\eqref{Eq:gammak}, and the summation runs over the magnetic Brillouin zone as before.
A $2\times 2$ Bogoliubov transformation yields two degenerate magnon modes with dispersion
\begin{align}
\omega_{\vec{k},1,2,}= 2d Js \sqrt{\lambda^2-\gamma_{\vec{k}}^2}\,.
\label{Eq:omegaxxzising}
\end{align}
At the $\SUtwo$-symmetric point at $\lambda=1$, these modes are gapless at $\vec k=0$. For small $k$ and small $(\lambda-1)$, the mode energy can be expanded as in Eq.~\eqref{Eq:Taylorexcitationenergyspinmodel}, but with a gap $\Delta_2\equiv\Delta$ given by
\begin{align}
\label{eq:delta2}
\Delta_2(\lambda)= 2Jsd \sqrt{\lambda^2-1}
\end{align}
and a velocity
\begin{align}
c_2(\lambda)= 2Js \sqrt{2d}\,.
\end{align}

As above, we evaluate the expressions \eqref{eq:cv} and \eqref{eq:al} in the continuum limit. For $T\gg\Delta$ we find
\begin{align}
\label{gammaxxz3}
&\lim_{\lambda \rightarrow 1^+} \Gamma=
\frac{4 d\zeta(d\!-\!1)}{(d\!+\!1)\zeta(d\!+\!1)} \Big(\frac{T}{Js}\Big)^{-2}
\end{align}
which is identical to the corresponding result in the XZ model up to a factor of two arising from the different $\Delta$ dependence of the gap, Eq.~\eqref{eq:delta1} vs. Eq.~\eqref{eq:delta2}.
For $T\ll\Delta$ we find
\begin{align}
\lim_{T\to 0} \Gamma(\lambda) = \frac{\lambda}{\lambda^2-1} ~\underset{\lambda\to1^+}{\to}~ \frac{1/2}{\lambda-1}
\end{align}
which agrees with the XZ-model result.

\subsubsection{XY Phase}

For $\lambda \leq 1$ the expansion is performed about a N\'eel state in the XY plane. The linear spin-wave Hamiltonian takes the form
\begin{align}
\Htwo
= 2d J s \sum_{\vec k} &\Big[ a^\dagger_{\vec{k}}a_{\vec{k}}+ b^\dagger_{\vec{k}}b_{\vec{k}} + \gamma_k (1-\lambda) (a^\dagger_{\vec{k}}b_{\vec{k}}+h.c.) \notag\\ &+ \gamma_k(1+\lambda) (a_{\vec{k}}b_{-\vec{k}}+h.c.) \Big]\,.
\end{align}
A $4\times4$ Bogoliubov transformation now yields two modes with dispersion
\begin{align}
&\omega_{\vec k,1}= 2d Js \sqrt{1+\gamma_{\vec{k}}\left(1-\lambda\left(1+\gamma_{\vec{k}}\right)\right)}, \nonumber \\
&\omega_{\vec k,2}= 2d Js \sqrt{1-\gamma_{\vec{k}}\left(1-\lambda\left(1-\gamma_{\vec{k}}\right)\right)},
\label{Eq.omega12xy}
\end{align}
which are non-degenerate except for $\lambda =1$ where $\omega_{\vec k,1}=\omega_{\vec k,2}$. For $\lambda<1$ the second mode represents the (in-plane) Goldstone mode of the XY phase whereas the first (out-of-plane) mode develops a gap, $\Delta_3\equiv\Delta$.
For small $k$ and small $(1-\lambda)$ a Taylor expansion of the mode energies yields
\begin{align}
&\omega_1 \approx \sqrt{\Delta_3^2+c_3^2|\vec{k}|^2}, \nonumber \\
&\omega_2 \approx c_4(\lambda)|\vec{k}|
\label{Eq:GoldstoneandgappedmodeenergyXYlambdaZ}
\end{align}
with
\begin{align}
&\Delta_3(\lambda)= 2d Js\sqrt{2(1-\lambda)}, \nonumber \\
&c_3(\lambda)= Js \sqrt{2d(3\lambda-1)}, \nonumber \\
&c_4(\lambda)= Js \sqrt{2d(1+\lambda)}.
\label{Eq.delta3}
\end{align}

Both specific heat and thermal expansion acquire contributions from both modes, such that the Gr\" uneisen ratio becomes
\begin{align}
\Gamma=\frac{\alpha_{gs}+\alpha_\Delta}{c_{gs}+c_{\Delta}}.
\label{Eq:gammagoldandgap}
\end{align}
In the regime $T\gg\Delta$ we recover the power laws $c_{gs}, c_\Delta \propto T^d$, moreover we have $\alpha_{gs}\propto T^d$ and $\alpha_\Delta \propto -T^{d-2}$. Evaluating the integrals yields
\begin{align}
\label{eq:gamxydiv}
&\lim_{\lambda\to 1^-} \Gamma=
-\frac{2 d \zeta(d\!-\!1)}{(d\!+\!1)\zeta(d\!+\!1)} \Big(\frac{T}{Js}\Big)^{-2}
\end{align}
with the sign being reversed compared to Eq.~\eqref{gammaxxz3}.
The opposite regime $T\ll\Delta$ contains a region where the contributions to $c_v$ and $\alpha$ from the gapped out-of-plane mode are negligible compared to that from the Goldstone mode. The latter scale as $T^d$ and are non-singular as $\lambda\to 1^-$ because $c_3(\lambda)$ is non-singular. This results in a positive and non-divergent Gr\" uneisen ratio
\begin{align}
\label{eq:gamxyconst}
\lim_{T\to 0} \Gamma(\lambda)
=\frac{1}{2(1+\lambda)} ~\underset{\lambda\to1^-}{\to}~ 1/4
\end{align}
in the region where the gapped mode can be neglected -- this does \emph{not} apply to the entire regime $T\ll\Delta$ as will become clear shortly.

To understand the crossover from the non-divergent behavior \eqref{eq:gamxyconst} to the divergent behavior \eqref{eq:gamxydiv}, we evaluate the gapped-mode contributions in the regime $T\ll\Delta$. By expanding the argument of the $\sinh$ in Eqs. \eqref{eq:cv} and \eqref{eq:al}  we find
\begin{align}
c_\Delta &= C_3(\lambda) (\beta\Delta)^{2+d/2} e^{-\beta\Delta}, \\
\alpha_\Delta &= - C_3(\lambda) (2d)^2 (\beta Js)^2 (\beta\Delta)^{d/2} e^{-\beta\Delta},
\end{align}
with the prefactor
\begin{align}
C_3(\lambda)= 2^{d/2-1} \tilde\Gamma(d/2) \Omega_d \frac{T^d}{ c_3(\lambda)^d}
\label{Eq:Cl}
\end{align}
where $\tilde\Gamma(x)$ is the Gamma function.
Hence, we can write $c_\Delta = T^d f(\beta\Delta)$ and $\alpha_\Delta = - T^{d-2} g(\beta\Delta)$ where both functions $f(x)$, $g(x)$ are exponentially suppressed with $x$. Recalling that $\alpha_{gs} \propto T^d$ we conclude that at {\em fixed} $\beta\Delta$ the gapped mode will dominate the thermal expansion at sufficiently low $T$, such that $\Gamma \propto -1/T^2$ upon approaching the QPT along trajectories of $\Delta/T=\rm const.$ even if $T\ll\Delta$.
The line separating non-divergent from divergent behavior of $\Gamma$ upon approaching the QPT has the form $\Delta/T \propto |\ln T/(Js)|$, obtained from solving $\alpha_{gs}=\alpha_\Delta$, as noted in the main text. Along this line the ratio $\Delta/T$ grows upon approaching the transition point.

%%%%%%%%%%%%%%%%%%%%%%%%%%%%%%%%%%%%%%%%%%%%%%%%%%%%%%%%%%%%%%%%%%%%%%%%%%%%%%%


\begin{thebibliography}{99}
\bibitem{gruendef}
In the literature slightly varying definitions of the Gr\"uneisen ratio (or Gr\"uneisen parameter) can be found. Since they only differ in prefactors which are non-singular at a QCP, they all share the property of being singular at continuous QPTs as well as symmetry-enhanced first-order QPTs.
\bibitem{sympicnote}
To make the Gr\"uneisen ratio of the XZ model \eqref{hxz} fully antisymmetric w.r.t. the transition point $\lambda=1$, $\Gamma(\lambda, T) = -\Gamma(1/\lambda, T)$, would require a re-scaling of the Hamiltonian by $1/\sqrt{\lambda}$ and a re-definition of the thermal expansion, i.e., by using $-\partial/\partial (1/\lambda)$ instead of $\partial/\partial\lambda$ for $\lambda<1$. For simplicity, we refrain from doing so, therefore the above relation only holds for the leading divergent term and for small $|\lambda-1|$.
\end{thebibliography}
\end{document}